\documentclass[aps,prl,floatfix]{revtex4-1}
\usepackage{graphicx}
\usepackage{amsmath}

\begin{document}
\title{A simple solver for the two-fluid plasma model based on PseudoSpectral Time-Domain algorithm}

 \author{B. Morel, R. Giust, K. Ardaneh and F. Courvoisier}
\email{ francois.courvoisier@femto-st.fr} 
\affiliation{ Institut FEMTO-ST, Universite Bourgogne Franche-Comte, 15B Avenue des Montboucons, 25030 Besancon cedex, France}

\begin{abstract}

We present a solver of 3D two-fluid plasma model for the simulation of short-pulse laser interactions with plasma. This solver resolves the equations of the two-fluid plasma model with ideal gas closure. We also include the Bhatnagar-Gross-Krook collision model. Our solver is based on PseudoSpectral Time-Domain (PSTD) method to solve Maxwell's curl equations. We use a Strang splitting to integrate Euler equations with source term: while Euler equations are solved with a composite scheme mixing Lax-Friedrichs and Lax-Wendroff schemes, the source term is integrated with a fourth-order Runge-Kutta scheme. This two-fluid plasma model solver is simple to implement because it only relies on finite difference schemes and Fast Fourier Transforms. It does not require spatially staggered grids. The solver was tested against several well-known problems of plasma physics. Numerical simulations gave results in excellent agreement with analytical solutions or with previous results from the literature.
\end{abstract}


\maketitle

\section{Introduction}
Since the early 1960s, the interaction of laser light with a plasma has revealed to be an extremely rich topic : laser-plasma accelerators, inertial fusion, X-Ray sources, nonlinear plasmonics or laser materials processing \cite{eliezer_applications_2008}. A number of plasma effects have been characterized in laser-plasma experiments, but many challenging problems remain \cite{kruer_physics_2003}. 
Simulations in laser-plasma interaction domain are essential for the understanding of the physical phenomena in high-intensity laser interactions. However, this requires efficient and precise numerical simulations. The hydrodynamic approach is particularly interesting to save computational effort when each of the species can be assumed in local thermodynamic equilibrium \cite{mckenna_laser-plasma_nodate}.

 The most fundamental approach for hydrodynamic code is the set of multi-fluid plasma equations \cite{gibbon_short_2005}. These equations can be derived by taking moments of the Vlasov equation with respect to velocity and truncating  moment equations by making additional assumptions. Here, we will consider a two-fluid system of electrons and ions. The closure of the system of equations is performed via ideal gas closure. This model describes the evolution in space and time of the density, mean velocity and pressure of electrons and ions fluids. The two-fluid plasma equations therefore consist of two sets of Euler equations with source term, as well as Maxwell's equations. More conventional plasma models, e.g., two-temperature plasma equations, single-fluid equations and MagnetoHydroDynamic (MHD) are derived from the two-fluid plasma model but require additional assumptions.

 \begin{figure}
\centering
  \includegraphics[scale=1.5]{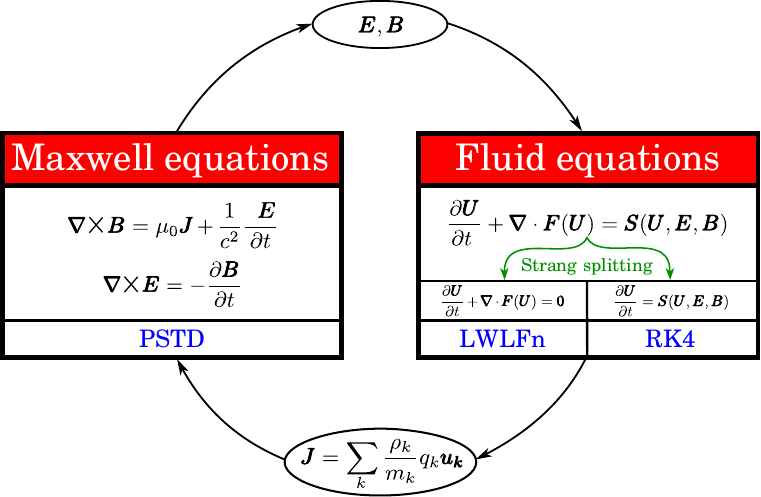} %
  \caption{General overview of the algorithm.}
  \label{fig:Fi1}
\end{figure}

Until now, only few numerical codes solve the complete two-fluid plasma equations \cite{abgrall_robust_2014}. However, their implementation is often complex for non-specialist groups. ANTHEM code was the first two-fluid plasma algorithm  \cite{mason_electromagnetic_1987,mason_hybrid_1986}. It is based on a flux-corrected transport (FCT) algorithm for fluids equations and a Finite-Difference Time Domain (FDTD) algorithm for the evolution of the fields. This code was used to simulate high-density plasmas out of reach to Particle-In-Cell (PIC) codes. It was although difficult to extend to arbitrary geometries because of the staggered grid of the FDTD scheme in the Yee algorithm \cite{yee_numerical_1966}.
More recently, U. Shumlak \textit{et al.} \cite{shumlak_approximate_2003} presented an algorithm based on Roe-type Riemann solver \cite{roe_approximate_1981} for the two-fluid plasma model. Time updates are carried out by treating the source term implicitly and the flux terms explicitly. This code was developed to observe two-fluid effects which are inaccessible to magnetohydrodynamic models. The same group added the high-order discontinuous Galerkin method to improve the result's accuracy \cite{loverich_discontinuous_2005,loverich_discontinuous_2011,srinivasan_analytical_2011, sousa_blended_2016}. A. Alvarez Laguna \textit{et al.} \cite{alvarez_laguna_fully-implicit_2018} introduce another algorithm where the two-fluid plasma equations are solved with a fully-implicit finite volume method for unstructured meshes combined with a modified version of Rusanov scheme \cite{rusanov_calculation_1961} for Maxwell equations. However, these solvers are relying on complex mathematical methods requiring specialized skills. Other approaches have been developed relying on finite difference algorithm: S. Baboolal \cite{baboolal_two-scale_2007} reports a one-dimensional scheme for shock and soliton simulations. H. Kumar \textit{et al.}\cite{kumar_entropy_2012} describe a scheme combining entropy conservative fluxes and suitable numerical diffusion operators. However, the first is limited to one dimension and the second requires heavy finite difference scheme to avoid entropy dissipation for shocks.

These codes are usually developed for complex problems as nuclear fusion and astrophysics, and also tested on very complex problems as the Brio and Wu shock tube test \cite{brio_upwind_1988}. Our code is tested with laser/plasma interaction problems with intensities around $10^{14}$ to $10^{15}$~W/cm$^2$, since this code is intended for the study of electron/hole plasma dynamics in solids. 

In this framework, we present a new simple method to solve 
equations of the adiabatic inviscid five-moment two-fluid plasma model with collisions. It only relies on finite-difference schemes and Fast Fourier Transforms (FFT) over a spatially non-staggered grid. This solver is intended to be used in absence of discontinuities.
Our approach is summarised in Fig.~\ref{fig:Fi1}. Maxwell's equations are solved using PseudoSpectral Time-Domain (PSTD) algorithm. The electromagnetic fields are passed to the fluid equations as source terms. We solve the fluid equations using a Strang splitting. In the splitting, the Euler equations (without source terms) are solved with a composite scheme, mixing Lax-Friedrichs and Lax-Wendroff schemes, while the source terms are integrated via fourth order Runge-Kutta scheme (RK4). The computed currents are then used as sources for Maxwell's equations.

This method has successfully been tested against classical plasma physics problems covering: i/ the three wave propagation modes in warm, fully ionized, isotropic plasmas; ii/ mode conversion and iii/ a  nonlinear laser-plasma interaction regime.

This paper is organised as follows. In sections 2, 3, 4 and 5 we briefly recall the models and methods on which our approach is based: the two-fluid plasma model, the collision model, the PseudoSpectral Time-Domain (PSTD) algorithm and the fluid equation solver. This will provide the reader a consistent way of writing the equations. Then, in section 6 we describe our algorithm and the stability conditions. In section 7, we present the 5 different tests validating the solver.

\section{Two-fluid plasma equations}

As for any hydrodynamic code, we start from the two-fluid plasma equations \cite{gibbon_short_2005}. We consider here a plasma composed of an electron fluid and an ion fluid where we assume no heat flow and no viscosity. The inviscid five-moment two-fluid plasma model equations can be derived from the Vlasov-Maxwell equations \cite{a._hakim_high_2006}. The two-fluid plasma model equations listed below consists of Euler equations with source term for each fluid, as well as Maxwell's equations. We first write the collisionless equations and we will see in the next section how they can be introduced as an additional source term. Following Ref.~\cite{a._hakim_high_2006}, the fluid equations are written as balance laws, {\it i.e.} a conservative form on the left side and a source term on the right one. This approach offers access to a large number of numerical methods.
\begin{equation}
\frac{\partial }{\partial t} {\bf U}  + {\bf \nabla} \cdot \left[ {\bf F}({\bf U})\right] = {\bf S}({\bf U},{\bf E},{\bf B})  
\label{eq:Econs}
\end{equation}
where ${\bf U}$ is the fluid variables vector:
\begin{equation}
{\bf U} \equiv \left[ \begin{matrix}
\rho_\mathrm{e} \\
\rho_\mathrm{e} {\bf u_\mathrm{e}} \\
\epsilon_\mathrm{e} \\
\rho_\mathrm{i} \\
\rho_\mathrm{i} {\bf u_\mathrm{i}} \\
\epsilon_\mathrm{i} \\
\end{matrix} \right] 
\end{equation}
${\bf F}({\bf U})$ is the flux tensor:
\begin{equation}
{\bf F}({\bf U}) \equiv  \left[ \begin{matrix}
\rho_\mathrm{i} {\bf u_\mathrm{e}} \\
\rho_\mathrm{e} {\bf u_\mathrm{e}} \otimes {\bf u_\mathrm{e}} + p_\mathrm{e} {\bf I}   \\
(\epsilon_\mathrm{e} + p_\mathrm{e}) {\bf u_\mathrm{e}} \\
\rho_\mathrm{i} {\bf u_\mathrm{i}} \\
\rho_\mathrm{i} {\bf u_\mathrm{i}} \otimes {\bf u_\mathrm{i}} + p_\mathrm{i} {\bf I} \\
(\epsilon_\mathrm{i} + p_\mathrm{i}) {\bf u_\mathrm{i}}
\end{matrix} \right]
\end{equation}
and ${\bf S}({\bf U},{\bf E},{\bf B})$ is the source term:
\begin{equation}
{\bf S}({\bf U},{\bf E},{\bf B})   \equiv \left[ \begin{matrix}
0 \\
\frac{\rho_\mathrm{e} q_\mathrm{e}}{m_\mathrm{e}} ({\bf E} + {\bf u_\mathrm{e}} {\bf \times} {\bf B} ) \\
\frac{\rho_\mathrm{e} q_\mathrm{e}}{m_\mathrm{e}} {\bf u_\mathrm{e}} \cdot {\bf E}  \\
0 \\
\frac{\rho_\mathrm{i} q_\mathrm{i}}{m_\mathrm{i}} ({\bf E} + {\bf u_\mathrm{i}} {\bf \times} {\bf B} ) \\
\frac{\rho_\mathrm{i} q_\mathrm{i}}{m_\mathrm{i}} {\bf u_\mathrm{i}} \cdot {\bf E} 
\end{matrix}  \right]
\end{equation}
Here, $e$ and $i$ are indexes related respectively to the electron fluid and to the ion fluid. ${\bf I}$ is the identity matrix and $\otimes$ is tensor product. $q_\mathrm{k}$ is the $k$ specie charge, $m_\mathrm{k}$ the specie mass, $\rho_\mathrm{k}$ the mass density, ${\bf u_\mathrm{k}}$ the mean velocity, $p_\mathrm{k}$ the pressure, $\epsilon_\mathrm{k}$ the fluid energy density, ${\bf E}$  the electric field and ${\bf B}$ the magnetic field. The system of equations (\ref{eq:Econs}) corresponds to continuity equations, motion equations and energy transport equations for electron and ion fluids.
The system of equations is closed with ideal gas closure for each fluid $k$ \cite{a._hakim_high_2006} :
\begin{equation}
\epsilon_\mathrm{k} \equiv \underbrace{\frac{p_\mathrm{k}}{\gamma -1 }}_{\textnormal{Thermal energy}} + \underbrace{\frac{1}{2}\rho_\mathrm{k} ||{\bf u_\mathrm{k}}||^2}_{\textnormal{Kinetic energy}}
\end{equation}
where $\gamma$ is the adiabatic index.

 The electric and magnetic fields appearing in the source term of the Euler equations are determined by the Maxwell equations:
\begin{equation}
{\bf \nabla} \cdot \epsilon_\mathrm{r} {\bf  E} = \frac{1}{\epsilon_\mathrm{0}} \underbrace{\left[ \frac{q_\mathrm{e}}{m_\mathrm{e}}\rho_\mathrm{e} + \frac{q_\mathrm{i}}{m_\mathrm{i}}\rho_\mathrm{i} \right]}_{\textnormal{Total charge density}}
\label{eq:MG}
\end{equation}
\begin{equation}
{\bf \nabla} \cdot {\bf B} = 0
\label{eq:M2}
\end{equation}
\begin{equation}
{\bf \nabla} {\bf \times} {\bf E} = - \frac{\partial {\bf B}}{\partial t}
\label{eq:curl1}
\end{equation}
 \begin{equation}
{\bf \nabla} {\bf \times} {\bf B} = \mu_\mathrm{0}  \underbrace{\left[ \frac{q_\mathrm{e}}{m_\mathrm{e}}\rho_\mathrm{e} {\bf u_\mathrm{e}} + \frac{q_\mathrm{i}}{m_\mathrm{i}}\rho_\mathrm{i} {\bf u_\mathrm{i}} \right]}_{\textnormal{Current density {\bf J}}}  + \frac{1}{c^2}\frac{\partial \epsilon_\mathrm{r} {\bf E}}{\partial t}
\label{eq:MA}
\end{equation}
where $\epsilon_\mathrm{0}$ and $\mu_\mathrm{0}$ are respectively the vacuum permittivity and permeability. $\epsilon_\mathrm{r}$ is the relative permittivity of the medium, and $c=(\epsilon_\mathrm{0} \mu_\mathrm{0})^{-1/2}$ is the speed of light.

\section{Collisions in the two-fluid plasma model}

The addition of a basic collisional model is straightforward in the two-fluid plasma model. One of the simplest model is the BGK (Bhatnagar-Gross-Krook) model \cite{Krook_1954}. In this model, it is assumed that the effect of collisions is to restore a situation of local equilibrium. It assumes that the local equilibrium is reached exponentially with time. The inclusion of the BGK collision model in the two-fluid plasma equations is handled through an additional source term given by \cite{Golant} :
\begin{equation}
{\bf S^{coll}}({\bf U})   \equiv \left[ \begin{matrix}
0 \\
 - \frac{\rho_\mathrm{e}}{m_\mathrm{e}} \mu \nu_\mathrm{ei} ( {\bf u_\mathrm{e}} - {\bf u_\mathrm{i}} ) \\
 - \kappa \nu_\mathrm{ei} \left[ \frac{p_\mathrm{e}}{\gamma -1} - \frac{\rho_\mathrm{e} m_\mathrm{i}}{\rho_\mathrm{i} m_\mathrm{e}}\frac{p_\mathrm{i}}{\gamma -1}   +  \frac{ \rho_\mathrm{e} }{2m_\mathrm{e}} \left(m_\mathrm{e} {\bf u_\mathrm{e}}^2 - m_\mathrm{i}{\bf u_\mathrm{i}}^2 + (m_\mathrm{i} -m_\mathrm{e}){\bf u_\mathrm{e}}\cdot{\bf u_\mathrm{i}}  \right)   \right]
 \\
0 \\
 - \frac{\rho_\mathrm{e}}{m_\mathrm{e}} \mu  \nu_\mathrm{ei} ( {\bf u_\mathrm{i}} - {\bf u_\mathrm{e}} )\\
  + \kappa \nu_\mathrm{ei} \left[ \frac{p_\mathrm{e}}{\gamma -1} - \frac{\rho_\mathrm{e} m_\mathrm{i}}{\rho_\mathrm{i} m_\mathrm{e}}\frac{p_\mathrm{i}}{\gamma -1}   +  \frac{ \rho_\mathrm{e} }{2m_\mathrm{e}} \left(m_\mathrm{e} {\bf u_\mathrm{e}}^2 - m_\mathrm{i}{\bf u_\mathrm{i}}^2 + (m_\mathrm{i} -m_\mathrm{e}){\bf u_\mathrm{e}}\cdot{\bf u_\mathrm{i}}  \right)   \right]
\end{matrix} \right]
\end{equation}
where $\mu = \frac{m_\mathrm{e} m_\mathrm{i}}{m_\mathrm{e} +m_\mathrm{i}}$ is the reduced mass, $\kappa = \frac{2 \mu}{m_\mathrm{e}+m_\mathrm{i}}$ the energy transfer coefficient, and $\nu_\mathrm{ei}$ the collision frequency for momentum transfer between electrons and ions. This collision term preserves momentum and energy conservation.

\section{The Maxwell solver}

 Numerical methods to solve Maxwell equations have been widely studied \cite{liu_pstd_1997,yee_numerical_1966,vincenti_ultrahigh-order_2018,shang_comparative_1996,munz_three-dimensional_2000,monorchio_hybrid_2004}. The PSTD algorithm \cite{liu_pstd_1997} is a pseudo-spectral method which is simple to implement and does not need a spatially staggered grid, in contrast with the Yee FDTD algorithm \cite{yee_numerical_1966}. The PSTD is therefore more flexible to be coupled with another algorithm without too many interpolations. Importantly, PSTD is low-dispersive since it provides accurate dispersion relation with only two cells per wavelength. The relative error for the PSTD algorithm with 2 cells per wavelength is smaller than that for the FDTD method with 16 cells per wavelength \cite{liu_pstd_1997}.
  The PSTD algorithm imposes periodicity because it is based on the Fast Fourier Transform (FFT), but this constraint can be removed by using Berenger's perfectly matched layers (PML) \cite{berenger_perfectly_1994}. Furthermore, the PSTD scheme allows to implement non-uniform background permittivity $\epsilon_\mathrm{r}$. It is also possible to model non-linear polarization term as shown in T.-W. Lee \textit{et al.} \cite{Lee_2004}. Therefore, PSTD algorithm is an excellent candidate to construct a simple and reliable hydrodynamic solver. 
  
We note that solving fluid equations and Maxwell's curl equations enforces the conservation of divergence properties of the fields \cite{bittencourt_fundamentals_2013}. Hence, if fields divergence properties are satisfied at an initial time, they will be satisfied at a later time. It is therefore not necessary to solve equations (\ref{eq:MG}) and (\ref{eq:M2}).

In PSTD, spatial derivatives are approximated using an FFT algorithm \cite{liu_pstd_1997}. Moreover, central finite difference is used for temporal derivatives. Thus, the spatial grid is unstaggered but the temporal grid is staggered. For example, if ${\bf E}$ is defined at $t=n\Delta t$, ${\bf B}$ is defined at $t=\left(n+ \frac{1}{2} \right) \Delta t$, where $\Delta t$ is the time step of the PSTD algorithm.

We present here 3D PSTD algorithm valid for a uniform medium, to which current density is added. The spatial derivatives are performed in Fourier space. Eq.~(\ref{eq:DerivaE}) shows the partial derivative of $E_\mathrm{y}$ with respect to $x$:
\begin{equation}
\left( \frac{\partial E_\mathrm{y}}{\partial x} \right)^\mathrm{n} = \textnormal{iFFT}_\mathrm{x} \left[i k_\mathrm{x} \textnormal{FFT}_\mathrm{x} \left[E_\mathrm{y}^\mathrm{n}\right]\right]
\label{eq:DerivaE}
\end{equation}
where $\textnormal{iFFT}_\mathrm{x}$ denotes to the inverse FFT along the $x$-axis, and $k_\mathrm{x}$ is the spatial frequency along the $x$-axis. All the spatial derivatives will be done in Fourier space as shown in Eq.~(\ref{eq:DerivaE}).

Knowing the magnetic field at $t^\mathrm{n-1/2}$ and spatial derivatives of electric field at $t^\mathrm{n}$, magnetic field ${\bf B}^\mathrm{n+1/2}$ can be calculated: 
\begin{equation}
B_\mathrm{x}^\mathrm{n+1/2} = B_\mathrm{x}^\mathrm{n-1/2} + \Delta t\left[ \left( \frac{\partial E_\mathrm{y}}{\partial z} \right)^\mathrm{n} -\left( \frac{\partial E_\mathrm{z}}{\partial y} \right)^\mathrm{n} \right]
\end{equation}
\begin{equation}
B_\mathrm{y}^\mathrm{n+1/2} = B_\mathrm{y}^\mathrm{n-1/2} + \Delta t \left[  \left( \frac{\partial E_\mathrm{z}}{\partial x} \right)^\mathrm{n} - \left( \frac{\partial E_\mathrm{x}}{\partial z} \right)^\mathrm{n}\right]
\end{equation}
\begin{equation}
B_\mathrm{z}^\mathrm{n+1/2} = B_\mathrm{z}^\mathrm{n-1/2} + \Delta t \left[  \left( \frac{\partial E_\mathrm{x}}{\partial y} \right)^\mathrm{n}  - \left( \frac{\partial E_\mathrm{y}}{\partial x} \right)^\mathrm{n}  \right]
\end{equation}
The spatial derivatives of ${\bf B}^\mathrm{n+1/2}$ are carried out as the same way than for electric field (see Eq.~(\ref{eq:DerivaE})).
Finally, ${\bf E}^\mathrm{n+1}$ is updated using ${\bf E}^\mathrm{n}$ and spatial derivatives of the magnetic field: 
\begin{equation}
E_\mathrm{x}^\mathrm{n+1} = E_\mathrm{x}^\mathrm{n} + c^2 \Delta t \left[ \left( \frac{\partial B_\mathrm{z}}{\partial y} \right)^\mathrm{n+1/2} - \left( \frac{\partial B_\mathrm{y}}{\partial z} \right)^\mathrm{n+1/2} \right] - \frac{\Delta t}{\epsilon_\mathrm{0}} J_\mathrm{x}
\end{equation}
\begin{equation}
E_\mathrm{y}^\mathrm{n+1} = E_\mathrm{y}^\mathrm{n} + c^2 \Delta t \left[\left( \frac{\partial B_\mathrm{x}}{\partial z} \right)^\mathrm{n+1/2} - \left( \frac{\partial B_\mathrm{z}}{\partial x} \right)^\mathrm{n+1/2} \right] - \frac{\Delta t}{\epsilon_\mathrm{0}} J_\mathrm{y}
\end{equation}
\begin{equation}
E_\mathrm{z}^\mathrm{n+1} = E_\mathrm{z}^\mathrm{n} + c^2 \Delta t
\left[  \left( \frac{\partial B_\mathrm{y}}{\partial x} \right)^\mathrm{n+1/2}  - \left( \frac{\partial B_\mathrm{x}}{\partial y} \right)^\mathrm{n+1/2}  \right]  - \frac{\Delta t}{\epsilon_\mathrm{0}} J_\mathrm{z}
\end{equation}
where $J_\mathrm{x}$, $J_\mathrm{y}$ and $J_\mathrm{z}$ are the current density. Current density will be provided later on from fluid variables. 

We finish this section by recalling the PSTD stability criterion in $D$ dimensions \cite{liu_pstd_1997} :
\begin{equation}
\Delta t \leq \frac{2}{\pi} \frac{\Delta x}{c\sqrt{D}  } 
\label{eq:stab}
\end{equation}
where $\Delta x$ is the spatial-step.  

\section{The fluid solver}

In this section, we first discuss the numerical method chosen for homogeneous 3D Euler equations. We then introduce the method used to integrate the source term.

\subsection{Numerical resolution of Euler equations}

In the vectorial Eq.~(\ref{eq:Econs}), fluid equations are written under conservative form. The conservative form of Euler equations emphasizes that mass density, momentum and energy are conserved. Computationally, there are some advantages in using the conservative form and consequently it gives access to a large class of numerical methods called conservative methods \cite{toro_riemann_2013}. Here, the goal is to numerically solve Eq.~(\ref{eq:Econs}) with a simple and reliable scheme.

As mentioned in the introduction, extremely reliable schemes have been developed in the past, that can even handle large discontinuities and shocks, but are excessively complex for problems without such strong requirements.
Very attractive schemes are the {\it composite} ones introduced by Liska and Wendroff \cite{liska_composite_1998}. Theses schemes are simple and high-resolution \cite{hagen_how_2007}. They consist of a temporal composition of basic classical finite difference schemes. Here we have chosen a composite scheme based on two-step Lax-Wendroff and two-step Lax-Friedrichs schemes.

\subsection{The LWLFn composite scheme}

The Euler equations are numerically solved using an LWLFn composite scheme \cite{liska_composite_1998}. It consists of applying $(n-1)$ times  the two-step Lax-Wendroff (LW) scheme and, afterwards, once the two-step Lax-Friedrichs (LF) scheme. The LF scheme is first order in space and is dissipative. In contrast, the LW scheme is second-order in space and is dispersive; this produces oscillations in the vicinity of strong gradient regions. Therefore, the composition of the two schemes is an effective way to overcome these numerical artifacts \cite{liska_composite_1999}. This approach avoids the need for an artificial viscosity term \cite{liska_composite_1998} or for the use of hybrid scheme which become complex in multidimensional problems \cite{kucharik_optimally-stable_2006}.

The parameter $n$ can be tuned depending on the problem. If the solution contains strong gradients, it is better to use a small number $n$ (typically LWLF4). On the other hand, when the solution is smooth, it is possible to use exclusively the two-step LW scheme. When shape of solution is initially completely unknown, it is advisable to begin numerical simulation with a small number $n$ and to increase it progressively. The LWLFn scheme offers a wider field of applications than LW or LF scheme alone, without additional complexity.

\subsubsection{Two-step Lax-Friedrichs scheme}

The classical (single-step) LF scheme is a basic method to solve Hyperbolic Partial Differential Equations (HPDE) \cite{leveque_finite_2002}. The 
two-step LF is a variant of the classical LF scheme which allows better damping of high-frequencies, and less damping of low and middle frequencies than the classical method \cite{shampine_two-step_2005}. It is a better method to filter high-frequency oscillations generate by the two-step LW scheme. This scheme is stable under CFL (Courant-Friedrichs-Lewy) conditions \cite{shampine_two-step_2005} but, its use alone is often limited because it is only first-order accurate. 

The two-step LF can be extended to multidimensional in a simple way with the symmetrized dimensionally-split scheme (SYS) technique \cite{kucharik_optimally-stable_2006}. The dimensionally-split schemes are created by successively approximating solutions of the 3D conservation laws in solutions of three 1D problems: $\frac{\partial }{\partial t} {\bf U} + \frac{\partial}{\partial x} {\bf F_x}({\bf U})= {\bf 0}$, $\frac{\partial }{\partial t} {\bf U} + \frac{\partial}{\partial y} {\bf F_y}({\bf U})= {\bf 0}$ and $\frac{\partial }{\partial t} {\bf U} + \frac{\partial}{\partial z} {\bf F_z}({\bf U})= {\bf 0}$. 

First, we define $L^\mathrm{x}$ the operator for the two-step LF step along $x$ :
\begin{eqnarray}
L^\mathrm{x} ({\bf U}^\mathrm{n}_\mathrm{j,l,m} ) =  \frac{1}{2} \left[ {\bf U}^\mathrm{n+1/2}_\mathrm{j+1/2,l,m}  + {\bf U}^\mathrm{n+1/2}_\mathrm{j-1/2,l,m} \right] \nonumber \\ - \frac{\Delta t}{ \Delta x} \left[ \right. {\bf F_x}^\mathrm{n+1/2}_\mathrm{j+1/2,l,m} -& {\bf F_x}^\mathrm{n+1/2}_\mathrm{j-1/2,l,m} \left. \right] 
\label{eq:SYS1}
\end{eqnarray}
with
\begin{equation}
{\bf U}^\mathrm{n+1/2}_\mathrm{j+1/2,l,m} = \frac{1}{2} \left[ {\bf U}^\mathrm{n}_\mathrm{j+1,l,m} + {\bf U}^\mathrm{n}_\mathrm{j,l,m} \right] - \frac{\Delta t}{2 \Delta x} \left[ {\bf F_x}^\mathrm{n}_\mathrm{j+1,l,m} - {\bf F_x}^\mathrm{n}_\mathrm{j,l,m} \right]
\end{equation}
\begin{equation}
{\bf U}^\mathrm{n+1/2}_\mathrm{j-1/2,l,m} = \frac{1}{2} \left[ {\bf U}^\mathrm{n}_\mathrm{j,l,m} + {\bf U}^\mathrm{n}_\mathrm{j-1,l,m} \right] - \frac{\Delta t}{2 \Delta x} \left[ {\bf F_x}^\mathrm{n}_\mathrm{j,l,m} - {\bf F_x}^\mathrm{n}_\mathrm{j-1,l,m} \right]
\end{equation}
where $j$, $l$ and $m$ are respectively indexes for $x$, $y$ and $z$ positions. To make the notation less cluttered, we have set ${\bf F_x}^\mathrm{n}_\mathrm{j,l,m} \equiv {\bf F_x}\left( {\bf U}_\mathrm{j,l,m}^\mathrm{n}\right)$.

Similarly, we define $L^\mathrm{y}$ the operator for the two-step LF step along $y$ :
\begin{eqnarray}
L^\mathrm{y} ({\bf U}^\mathrm{n}_\mathrm{j,l,m}) = \frac{1}{2} \left[ {\bf U}^\mathrm{n+1/2}_\mathrm{j,l+1/2,m} + {\bf U}^\mathrm{n+1/2}_\mathrm{j,l-1/2,m} \right] \nonumber \\  - \frac{\Delta t}{ \Delta y} \left[ \right.{\bf F_y}^\mathrm{n+1/2}_\mathrm{j,l+1/2,m} -& {\bf F_y}^\mathrm{n+1/2}_\mathrm{j,l-1/2,m} \left. \right] 
\label{eq:SYS2}
\end{eqnarray}
with
\begin{equation}
{\bf U}^\mathrm{n+1/2}_\mathrm{j,l+1/2,m} = \frac{1}{2} \left[ {\bf U}^\mathrm{n}_\mathrm{j,l+1,m} + {\bf U}^\mathrm{n}_\mathrm{j,l,m} \right] - \frac{\Delta t}{2 \Delta y} \left[ {\bf F_y}^\mathrm{n}_\mathrm{j,l+1,m} - {\bf F_y}^\mathrm{n}_\mathrm{j,l,m} \right]
\end{equation}
\begin{equation}
{\bf U}^\mathrm{n+1/2}_\mathrm{j,l-1/2,m} = \frac{1}{2} \left[ {\bf U}^\mathrm{n}_\mathrm{j,l,m} + {\bf U}^\mathrm{n}_\mathrm{j,l-1,m} \right] - \frac{\Delta t}{2 \Delta y} \left[ {\bf F_y}^\mathrm{n}_\mathrm{j,l,m} - {\bf F_y}^\mathrm{n}_\mathrm{j,l-1,m} \right]
\end{equation}
Similarly, we define $L^\mathrm{z}$ the operator for the two-step LF step along $z$ :
\begin{eqnarray}
L^\mathrm{z} ({\bf U}^\mathrm{n}_\mathrm{j,l,m}) = \frac{1}{2}\left[ {\bf U}^\mathrm{n+1/2}_\mathrm{j,l,m+1/2} + {\bf U}^\mathrm{n+1/2}_\mathrm{j,l,m-1/2} \right] \nonumber \\ - \frac{\Delta t}{ \Delta z} \left[ \right.  {\bf F_z}^\mathrm{n+1/2}_\mathrm{j,l,m+1/2} -& {\bf F_z}^\mathrm{n+1/2}_\mathrm{j,l,m-1/2} \left. \right] 
\label{eq:SYS3}
\end{eqnarray}
with
\begin{equation}
{\bf U}^\mathrm{n+1/2}_\mathrm{j,l,m+1/2} = \frac{1}{2} \left[ {\bf U}^\mathrm{n}_\mathrm{j,l,m+1} + {\bf U}^\mathrm{n}_\mathrm{j,l,m} \right] - \frac{\Delta t}{2 \Delta z} \left[ {\bf F_z}^\mathrm{n}_\mathrm{j,l,m+1} - {\bf F_z}^\mathrm{n}_\mathrm{j,l,m} \right]
\end{equation}
\begin{equation}
{\bf U}^\mathrm{n+1/2}_\mathrm{j,l,m-1/2} = \frac{1}{2} \left[ {\bf U}^\mathrm{n}_\mathrm{j,l,m} + {\bf U}^\mathrm{n}_\mathrm{j,l,m-1} \right] - \frac{\Delta t}{2 \Delta z} \left[ {\bf F_z}^\mathrm{n}_\mathrm{j,l,m} - {\bf F_z}^\mathrm{n}_\mathrm{j,l,m-1} \right]
\end{equation}
The symmetrized dimensionally-split scheme (SYS) allows to obtain the value ${\bf U}^\mathrm{n+1}_\mathrm{j,l,m}$ from ${\bf U}^\mathrm{n}_\mathrm{j,l,m}$ by averaging all possible simple dimensionally-split schemes \cite{kucharik_optimally-stable_2006}:
\begin{eqnarray}
{\bf U}^\mathrm{n+1}_\mathrm{j,l,m} = \frac{1}{6} ( L^\mathrm{x} L^\mathrm{y} L^\mathrm{z} + L^\mathrm{x} L^\mathrm{z} L^\mathrm{y} + L^\mathrm{y} L^\mathrm{x} L^\mathrm{z} \nonumber \\+ L^\mathrm{y} L^\mathrm{z} L^\mathrm{x} 
+ L^\mathrm{z} L^\mathrm{x} L^\mathrm{y} + L^\mathrm{z} &L^\mathrm{y} L^\mathrm{x} )  {\bf U}^\mathrm{n}_\mathrm{j,l,m}  
\label{eq:Symmetrization}
\end{eqnarray}

\subsubsection{Two-step Lax-Wendroff scheme}

The LW scheme is a classical numerical method for HPDE, based on finite differences \cite{lax_systems_1960}. But for non-linear HPDE, the classical Lax-Wendroff scheme requires evaluating Jacobian matrices. Richtmyer \cite{richtmyer_survey_1962} proposed a two-step LW scheme in order to avoid calculating Jacobian matrices. The stability of this scheme is constrained by CFL conditions \cite{richtmyer_survey_1962}.

The 3D two-step LW is obtained just by replacing equations (\ref{eq:SYS1}), (\ref{eq:SYS2}) and (\ref{eq:SYS3}) respectively by :
\begin{equation}
L^\mathrm{x} ({\bf U}^\mathrm{n}_\mathrm{j,l,m} ) =  {\bf U}^\mathrm{n}_\mathrm{j,l,m}  - \frac{\Delta t}{ \Delta x} \left[ {\bf F_x}^\mathrm{n+1/2}_\mathrm{j+1/2,l,m} - {\bf F_x}^\mathrm{n+1/2}_\mathrm{j-1/2,l,m} \right] 
\end{equation}
\begin{equation}
L^\mathrm{y} ({\bf U}^\mathrm{n}_\mathrm{j,l,m}) =  {\bf U}^\mathrm{n}_\mathrm{j,l,m}  - \frac{\Delta t}{ \Delta y} \left[ {\bf F_y}^\mathrm{n+1/2}_\mathrm{j,l+1/2,m} - {\bf F_y}^\mathrm{n+1/2}_\mathrm{j,l-1/2,m} \right] 
\end{equation}
and
\begin{equation}
L^\mathrm{z} ({\bf U}^\mathrm{n}_\mathrm{j,l,m}) =  {\bf U}^\mathrm{n}_\mathrm{j,l,m}  - \frac{\Delta t}{ \Delta z} \left[ {\bf F_z}^\mathrm{n+1/2}_\mathrm{j,l,m+1/2} - {\bf F_z}^\mathrm{n+1/2}_\mathrm{j,l,m-1/2} \right] 
\end{equation}

\subsection{Source term integration}

In this subsection, we select a numerical technique to integrate the source term in Eq.~(\ref{eq:Econs}), which is a serious numerical challenge according to \cite{toro_riemann_2013}. The most widely used method is the splitting method which splits the system into two sub-problems, homogeneous system and the source system. Each of these systems are solved separately. Here, we employ the Strang splitting technique as presented in \cite{strang_construction_1968} and \cite{a._hakim_high_2006}.
This splitting is second-order accurate whereas a classical Godunov splitting is only first-order accurate \cite{leveque_finite_2002}.
In the Strang splitting, the system (\ref{eq:Econs}) is split as follows:
\begin{equation}
\frac{\partial }{\partial t} {\bf U} + {\bf \nabla} \cdot \left[ {\bf F}({\bf U})\right] = {\bf 0}
\label{eq:H11}
\end{equation}
\begin{equation}
\frac{\partial }{\partial t} {\bf U} = {\bf S}({\bf U},{\bf E},{\bf B})  
\label{eq:H12}
\end{equation}

Eq.~(\ref{eq:H11}) is a HPDE system while Eq.~(\ref{eq:H12}) is a Ordinary Differential Equation (ODE) system. The HPDE system is solved using LWLFn scheme and a fourth-order Runge-Kutta (RK4) scheme \cite{butcher_numerical_2004} is used to integrate the ODE system. In the Strang splitting, we first solved Eq.~(\ref{eq:H12}) with an RK4 scheme over $\Delta t /2$, then the homogeneous Eq.~(\ref{eq:H11}) is integrated over $\Delta t$ with an LWLFn scheme, and finally Eq.~(\ref{eq:H12}) is again solved with an RK4 over time step $\Delta t/2$. 

\section{Solver structure}

As presented in previous sections, we solve Maxwell's curl equations with a PSTD algorithm and the combination of LWLFn and RK4 solves the fluid equations based on a Strang splitting. The aim of this section is to describe the coupling of these two algorithms. This coupling is performed via the current density for Maxwell's equations and via the electromagnetic fields for the the fluid equations.

\begin{figure*}
\centering
\includegraphics[scale=0.8]{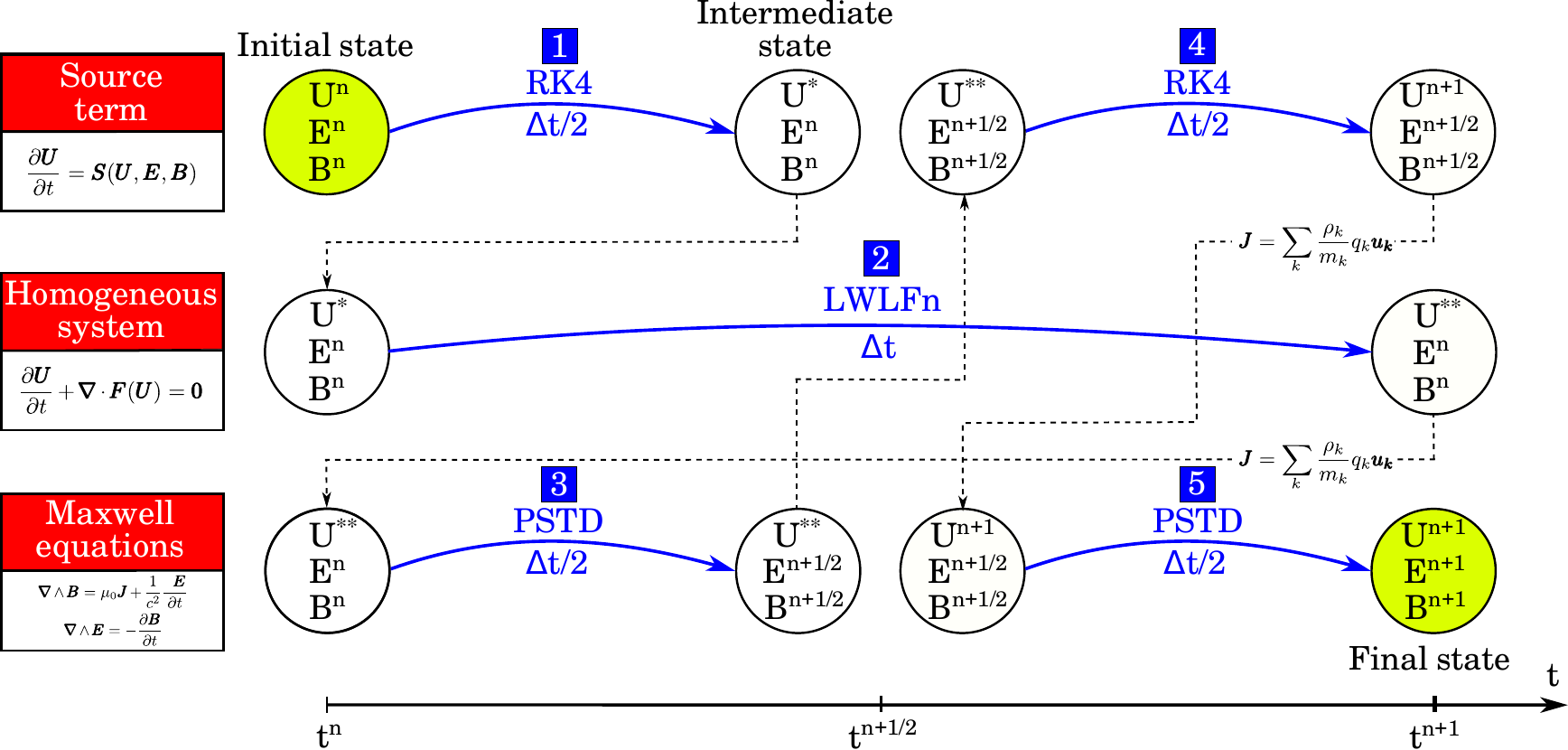}
  \caption{Description of the algorithm to solve the two-fluid plasma equations. An RK4 scheme is used to integrate the source term, an LWLFn scheme is used to integrate the homogeneous system and a PSTD algorithm is used to advance the fields. The algorithm involves five steps shown in blue to advance fields and fluid variables from a time step $n$ to $n+1$.}
  \label{fig:sch}
\end{figure*}

The full algorithm is detailed in Fig.~\ref{fig:sch}. This figure shows how to advance the values of fluid variables ${\bf U}$, electric field ${\bf E}$ and magnetic field ${\bf B}$ from time $t^\mathrm{n}$ to  $t^\mathrm{n+1}$. 

The initial state is defined as:
\begin{itemize}
\item fluid variables vector ${\bf U}^\mathrm{n}$ is known at time $t^\mathrm{n}$. 
\item The electric field is known at time $t^\mathrm{n}$. Due to the staggered temporal grid of PSTD, the magnetic field ${\bf B}$ is known at time $t^{\mathrm{\mathrm{\mathrm{n-1/4}}}}$ and at $t^{\mathrm{\mathrm{\mathrm{n+1/4}}}}$. We deduce ${\bf B^\mathrm{n}}$ with a linear interpolation between ${\bf B}^{\mathrm{n-1/4}}$ and ${\bf B}^{\mathrm{n+1/4}}$.
\end{itemize}

\hspace{-0.53cm}Then algorithm follows five main steps:
\begin{itemize} 
\item Step 1: Integrate the source term (\ref{eq:H12}) with an RK4 scheme over a temporal step $\Delta t/2$ by using ${\bf E}^\mathrm{n}$, ${\bf B}^\mathrm{n}$ and ${\bf U}^\mathrm{n}$ to obtain the intermediate value of fluid variables ${\bf U^*}$.
\item Step 2: Integrate the homogeneous system (\ref{eq:H11}) with an LWLFn scheme over a temporal step $\Delta t$ using fluid variables vector ${\bf U}^*$ to obtain a new intermediate value ${\bf U}^{**}$.
\item Step 3: Calculate the current density ${\bf J}$  with densities and velocities from ${\bf U}^{**}$.
Then, carry out a PSTD step with ${\bf J}$ to calculate ${\bf E}^\mathrm{n+1/2}$ and ${\bf B}^{\mathrm{\mathrm{\mathrm{n+3/4}}}}$. Calculate ${\bf B}^\mathrm{n+1/2}$ using a linear interpolation between ${\bf B}^{\mathrm{n+1/4}}$ and ${\bf B}^{\mathrm{n+3/4}}$.
\item Step 4: Integrate the source term (\ref{eq:H12}) with an RK4 algorithm over a temporal step $\Delta t/2$ using ${\bf U}^{**}$, ${\bf E}^\mathrm{n+1/2}$ and ${\bf B}^\mathrm{n+1/2}$ to obtain the final value of fluid variables ${\bf U}^\mathrm{n+1}$.
\item Step 5: Calculate the current density ${\bf J}$ with densities and velocities from ${\bf U}^\mathrm{n+1}$.
Then, carry out a PSTD step with ${\bf J}$ to calculate ${\bf E}^\mathrm{n+1}$ and ${\bf B}^{\mathrm{\mathrm{\mathrm{n+5/4}}}}$. Finally, calculate ${\bf B}^\mathrm{n+1}$ using a linear interpolation between ${\bf B}^{\mathrm{n+3/4}}$ and ${\bf B}^{\mathrm{n+5/4}}$.
\end{itemize} 

In order to capture the physics of the system, it is necessary to resolve the maximal frequency of the system \cite{loverich_discontinuous_2011}, which is either the laser pulsation $\omega_0$ or the hybrid pulsation \cite{a._hakim_high_2006}:

\begin{equation}
\omega_\mathrm{max} = \sqrt{ \omega^2_\mathrm{pe} + \omega^2_\mathrm{ce}} 
\label{omegh}
\end{equation}
where $\omega_\mathrm{pe} = \sqrt{\rho_\mathrm{e} q_\mathrm{e}^2 / m_\mathrm{e}^2 \epsilon_\mathrm{0}}$ is the electron plasma pulsation and $\omega_{ce} = q_\mathrm{e} B / m_\mathrm{e}$ is the cyclotron plasma pulsation.

In practice, when transverse electromagnetic waves are expected or injected, we use the stability criterion (Eq. (\ref{eq:stab})) to ensure the stability of the PSTD algorithm, which is usually the most demanding constraint. In absence of transverse electromagnetic waves, the sampling is adapted to resolve both the plasma thermal velocity and the hybrid pulsation.

\section{Validation of the solver}

The implementation of the LWLFn algorithm was first validated against the Noh test problem in 2D and 3D as in Refs.~\cite{liska_comparison_2003,kucharik_optimally-stable_2006}, which is known as a demanding test. We obtained quantitatively identical results. We note that for simpler tests such as the smooth problem of \cite{kucharik_optimally-stable_2006}, the symmetrization of Eq.~(\ref{eq:Symmetrization}) is not absolutely necessary. A basic spatial splitting such as ${\bf U}^\mathrm{n+1}_\mathrm{j,l,m} = L^\mathrm{x} L^\mathrm{y} L^\mathrm{z}  ({\bf U}^\mathrm{n}_\mathrm{j,l,m} ) $ is enough to obtain reliable results while significantly saving computation time.

Now, we validate our full solver against several classical plasma physics problems in which analytical solutions are available. In a warm fully ionized isotropic plasma, there are three modes of wave propagation \cite{bittencourt_fundamentals_2013}. The goal of the first two tests is to retrieve with the solver the dispersion relations of these three modes. In the third test, we check mode conversion between an electromagnetic mode and a longitudinal electron plasma mode. The fourth test validates the integration of the collisional model. Finally in the fifth test, we demonstrate that our solver is also valid in the nonlinear regime such as wakefield generation driven by ponderomotive force. We remark that in all the following tests, we have used the general framework of the LWLFn with n=5000 to keep the generality of our approach. Although in our case the LW scheme could be sufficient, problems with higher gradients may require to increase the number of Lax-Friedrichs steps in order to obtain convergence of the numerical solution (for example in the 3D Noh test of reference \cite{kucharik_optimally-stable_2006}, as we used to validate the fluid solver of our code). For instance, the dispersion curves shown in Fig.~\ref{fig:lang} are identical whether the pure LW or LWLF5000 are used.

The simulation setups and the boundary conditions (BC) for the five tests are summarized in table~\ref{tab:Tab1}. For all these tests, we set $m_\mathrm{e} = 9.11 \times 10^{-31}$~kg, $q_\mathrm{e} = -1.60  \times 10^{-19}$~C, $q_\mathrm{i} = -q_\mathrm{e}$, $\epsilon_\mathrm{0} = 8.85  \times 10^{-12}$~F/m, $\epsilon_\mathrm{r}=1$, $c=3.00  \times 10^8$~m/s and the Boltzmann constant is $k_\mathrm{B} = 1.38  \times 10^{-23}$~J/K. 

\begin{table*}
\caption{Summary of different variables for the five tests.}
\centering
\begin{tabular}{|p{3.25cm}||c|c|c|c|c|}
\hline
Parameter & Test 1 &  Test 2 &  Test 3 & Test 4 & Test 5  \\
\hline 
$\Delta x = \Delta y = \Delta z$ & $10$~nm &  $0.2$~nm &  $60$~nm & $20$~nm & $10$~nm  \\
\hline
$\Delta t$ & $14$~as &  $6.34$~as  & $96$~as  & $28$~as & $14$~as  \\
\hline
LWLFn & LWLF5000 &  LWLF5000 &  LWLF5000 & LWLF5000 & LWLF5000  \\
\hline
$N_\mathrm{x}$ & 4096 &  16384  & 256  & 2048 & 4096  \\
\hline
$N_\mathrm{y}$ & 2  & 2 &  2 & 2 &2  \\
\hline
$N_\mathrm{z}$ & 2   &  2 & 256 & 2 &2  \\
\hline
$x$ PSTD BC & Periodic  &  Periodic & PML & Periodic & Periodic \\
\hline
$y$ PSTD BC & Periodic &  Periodic &  Periodic & Periodic & Periodic \\
\hline
$z$ PSTD BC & PML  & PML & PML & PML & PML \\
\hline
$x$ LWLFn BC &  Periodic  &  Periodic & Open & Periodic & Periodic \\
\hline
$y$ LWLFn BC & Periodic & Periodic &  Periodic & Periodic & Periodic \\
\hline
$z$ LWLFn BC & Open  & Open &  Open & Open & Open \\
\hline
 Initial ${\bf u_\mathrm{e}}$ and ${\bf u_\mathrm{i}}$ & ${\bf 0}$ &  ${\bf 0}$ &  ${\bf 0}$ & ${\bf 0}$ & ${\bf 0}$\\
\hline
 $m_\mathrm{i}/m_\mathrm{e}$ & 1800 &  25  & 1800  & 1800 & 1800  \\
\hline
 Initial $p_\mathrm{e}$ and $p_\mathrm{i}$ & 0  & see Test 2 &  0 & 0 & 0  \\
\hline
 $\nu_\mathrm{ei}$ (fs$^{-1}$) & 0 & 0 &  0 & variable  &0.5 \\
\hline
 $\gamma$ & 5/3  & 5/3 &  5/3 & 5/3 & 5/3 \\
\hline
\end{tabular}
\label{tab:Tab1}
\end{table*}

\subsection{Test 1: Transverse electromagnetic wave propagation in an unmagnetized plasma}

The dispersion relation for transverse electromagnetic plasma mode in unmagnetized plasma is given by \cite{bittencourt_fundamentals_2013}:
\begin{equation}
\omega^2 = \omega^2_\mathrm{pe} +\omega^2_\mathrm{pi} +\left( \frac{2\pi}{\lambda_\mathrm{p}} \right)^2 c^2
\end{equation}
where $\omega_\mathrm{pe}$ and $\omega_\mathrm{pi}$ are respectively the electron and ion plasma frequencies.
Thus the wavelength of the electromagnetic wave in the plasma is:
\begin{equation}
\lambda_\mathrm{p}  = \frac{2\pi c}{\sqrt{\omega^2 - \omega^2_\mathrm{pe} - \omega^2_\mathrm{pi} }}
\label{eq:lp}
\end{equation}

For numerical simulations, we consider the following initial densities and fields:
\begin{itemize}
\item The initial densities are uniform. Several simulations are performed for different densities ranging between $0.05 \times 10^{21}$~cm$^{-3}$ and $1.4 \times 10^{21}$~cm$^{-3}$.
\item The electric and magnetic fields are initially set to zero everywhere. A monochromatic plane wave propagating in positive $z$-direction is injected with a transient on $\sim 50$ fs. The wave is polarized along the $x$-direction with a wavelength $\lambda = 0.8~\mu$m and an amplitude in vacuum $E_{0} = 10^{10}$ V/m. 
\end{itemize}

\begin{figure}
\centering
\includegraphics[scale=0.75]{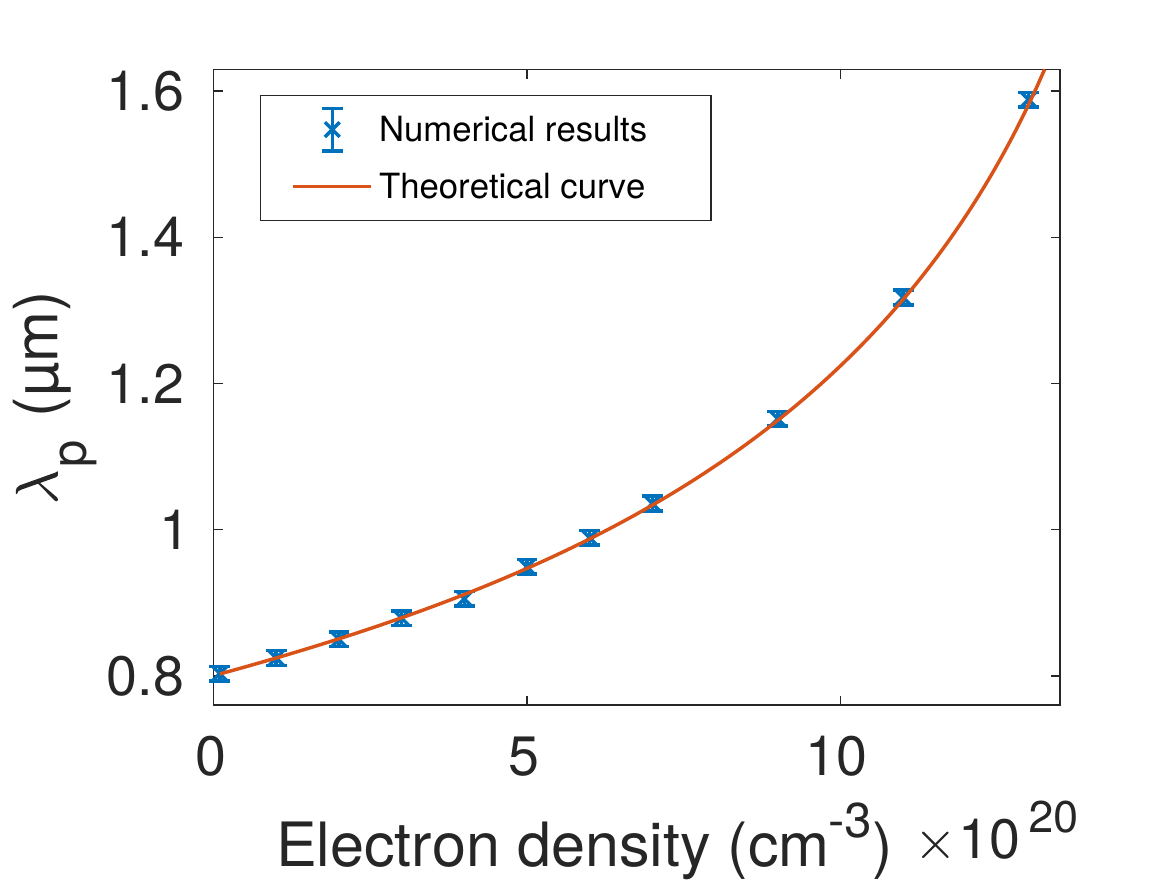}
\caption{Wavelength of the transverse electromagnetic waves in the plasma as function of electron density. Numerical results are shown in blue points while the red curve is plotted from the theoretical relation \ref{eq:lp}.}
\label{fig:long2}
\end{figure}

The final time of the simulations is $t_\mathrm{final}=150$~fs.
For each simulation, a $\textnormal{FFT}_\mathrm{z}$ is performed on $E_\mathrm{x}$ field to extract the wavelength in the plasma. The wavelengths from the numerical simulations are plotted as a function of plasma density with blue points in Fig.~\ref{fig:long2}. The error bar is defined by the spatial frequency sampling. The numerical results are in excellent agreement with the theoretical curve from Eq.~(\ref{eq:lp}).

\subsection{Test 2: Longitudinal modes in a warm plasma}

Isotropic plasmas support two longitudinal plasma waves: longitudinal electron plasma waves and longitudinal ion plasma waves. Linearization of the two-fluid plasma equations for small amplitude perturbations gives the following dispersion relation for longitudinal electron plasma mode \cite{bittencourt_fundamentals_2013}:
\begin{equation}
\omega^2 = \omega^2_\mathrm{pe} + \omega^2_\mathrm{pi} + \gamma \frac{k_\mathrm{B} T_\mathrm{e}}{m_\mathrm{e}} k^2
\label{eq:x1}
\end{equation}
where $T_\mathrm{e}$ is the electron temperature. 
 The dispersion relation of the longitudinal ion plasma wave in low-frequency limit $\left( \omega^2 \ll \omega^2_\mathrm{pi} \left[ 1 + \frac{T_\mathrm{i}}{T_\mathrm{e}} \right] \right)$ reads \cite{bittencourt_fundamentals_2013}:
\begin{equation}
\omega^2 = \gamma \frac{k_\mathrm{B} \left( T_\mathrm{i} +T_\mathrm{e} \right)}{m_\mathrm{i}}k^2
\label{eq:x2}
\end{equation}
where $T_\mathrm{i}$ is the ion temperature.
 In the high-frequency limit $\left(\omega^2 \gg \omega^2_\mathrm{pi} \left[ 1 + \frac{T_\mathrm{i}}{T_\mathrm{e}} \right]\right)$, the dispersion relation of the longitudinal ion plasma wave is given by \cite{bittencourt_fundamentals_2013}:
\begin{equation}
\omega^2 = \gamma \frac{k_\mathrm{B}  T_\mathrm{i}}{m_\mathrm{i}}k^2
\label{eq:x3}
\end{equation}
The goal of this second test is to reproduce relations (\ref{eq:x1}), (\ref{eq:x2}) and (\ref{eq:x3}). For this purpose, we consider the following initial conditions:
\begin{itemize}
\item The initial ion density is uniform: $\rho_\mathrm{i}/m_\mathrm{i} = 5 \times 10^{21}$~cm$^{-3}$.
\item  The initial electron density is uniform with a small perturbation in the vicinity of $z=0$~: \\ $\rho_\mathrm{e} = \rho_\mathrm{e0} \left[ 1 + \Delta d \frac{z}{W_\mathrm{z}} \exp \left( -z^2/W^2_\mathrm{z} \right) \right]$ where $\rho_{e0}/m_\mathrm{e} =  5 \times 10^{21}$~cm$^{-3}$, $\Delta d = 10^{-10}$ and $W_\mathrm{z} = 1$~nm. 
\item Based on Gauss's law, the $E_\mathrm{z}$ component associated to this perturbation is $E_\mathrm{z} = -\frac{\rho_\mathrm{e0} q_\mathrm{e} W_\mathrm{z} \Delta d}{2 m_\mathrm{e} \epsilon_\mathrm{0}} \exp \left(-z^2/W^2_\mathrm{z} \right)$. The others components of electric and magnetic fields are initially set to zero. 
\item The initial electron and ion temperatures are: $T_\mathrm{e} = 5 \times 10^6$~K and $T_\mathrm{i} = 2 \times 10^5$~K respectively. The initial pressures are linked to temperature via ideal gas law ($p=\frac{\rho}{m}k_B T$). 
\item The ion-to-electron mass ratio $m_\mathrm{i}/m_\mathrm{e}$ is fixed to 25 so as to observe both electron and ion waves in the same simulation.
\end{itemize}

\begin{figure}
\centering
\includegraphics[scale=0.85]{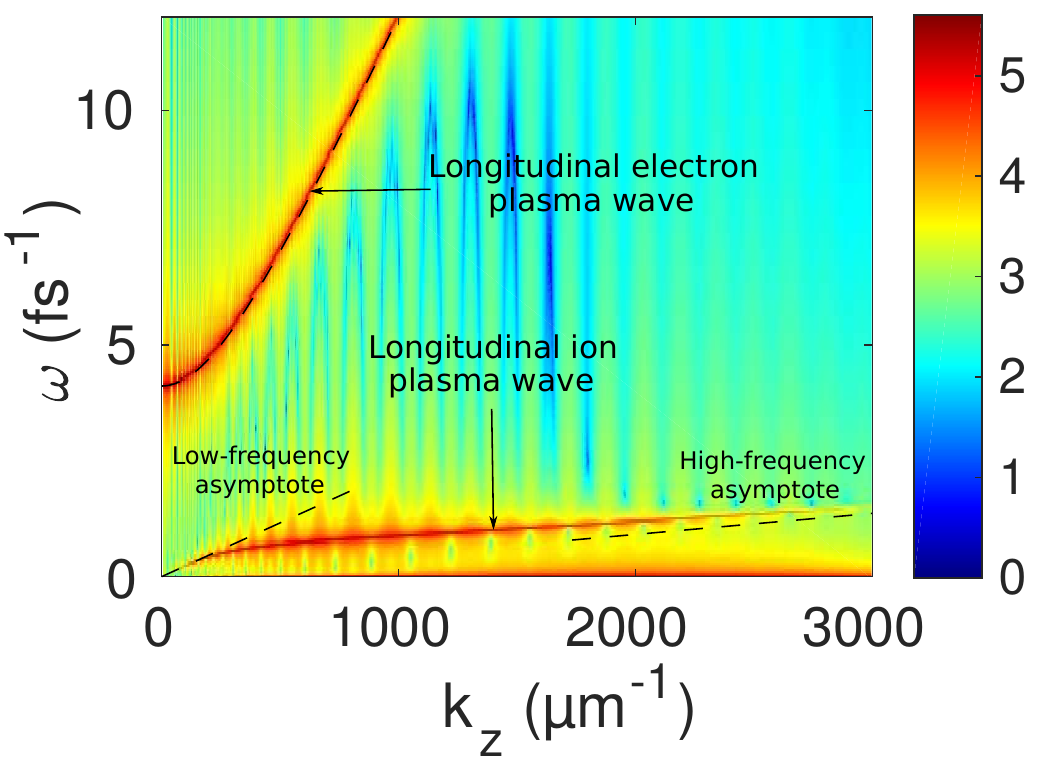}
\caption{Dispersion diagram ($\omega$,$k_z$) of longitudinal waves in a warm plasma. Black dashed lines correspond to the theoretical dispersion relations of longitudinal electron waves and longitudinal ion waves in two asymptotic limits (low and high frequency). The colorbar is in logarithmic scale.}
\label{fig:lang}
\end{figure}

We plot in log scale in Fig.~\ref{fig:lang} the dispersion diagram ($\omega$,$k_z$) of the $E_\mathrm{z}$ field after 130~fs of simulation. The dispersion relation of the longitudinal electron plasma wave (Eq. \ref{eq:x1}), and dispersion relation of the longitudinal ion plasma waves in two asymptotic limits (Eqs. (\ref{eq:x2}) and (\ref{eq:x3})), are shown as black dashed lines. The numerical results show a very good agreement with the analytical ones. We remark that in this case, since only small gradients are involved,  the first order LF scheme alone, the second order LW scheme alone or the composite scheme LWLFn provide the same result.

\subsection{Test 3: Wave conversion on plasma density ramp}
\label{sec:waveConversion}
Here we test wave conversion in a cold, unmagnetized, collisionless plasma. Wave conversion occurs when an electromagnetic wave, p-polarized, is obliquely incident onto a inhomogeneous plasma \cite{speziale_linear_1977}. At the critical surface, the local plasma frequency equals the frequency of the electromagnetic wave and since the incident electric field has a component along the density gradient, a plasma wave is driven resonantly. Therefore, part of the  energy of the incident light wave is transferred to the electron plasma wave. Analytical solutions to this difficult problem usually require a number of approximations. This field of research has been extensively investigated and we have selected out for comparison, the results from the literature considered as the most accurate. Speziale {\it et al.} described the asymptotic behaviors \cite{speziale_linear_1977}, Forslund {\it et al.} obtained numerical results based on PIC simulations \cite{forslund_theory_1975}, and the analytical results of Hinkel-Lipsker {\it et al.} match well with numerical simulations and experimental results \cite{hinkel-lipsker_analytic_1989,kruer_physics_2003}.

\begin{figure*}
\centering
\includegraphics[scale=0.62]{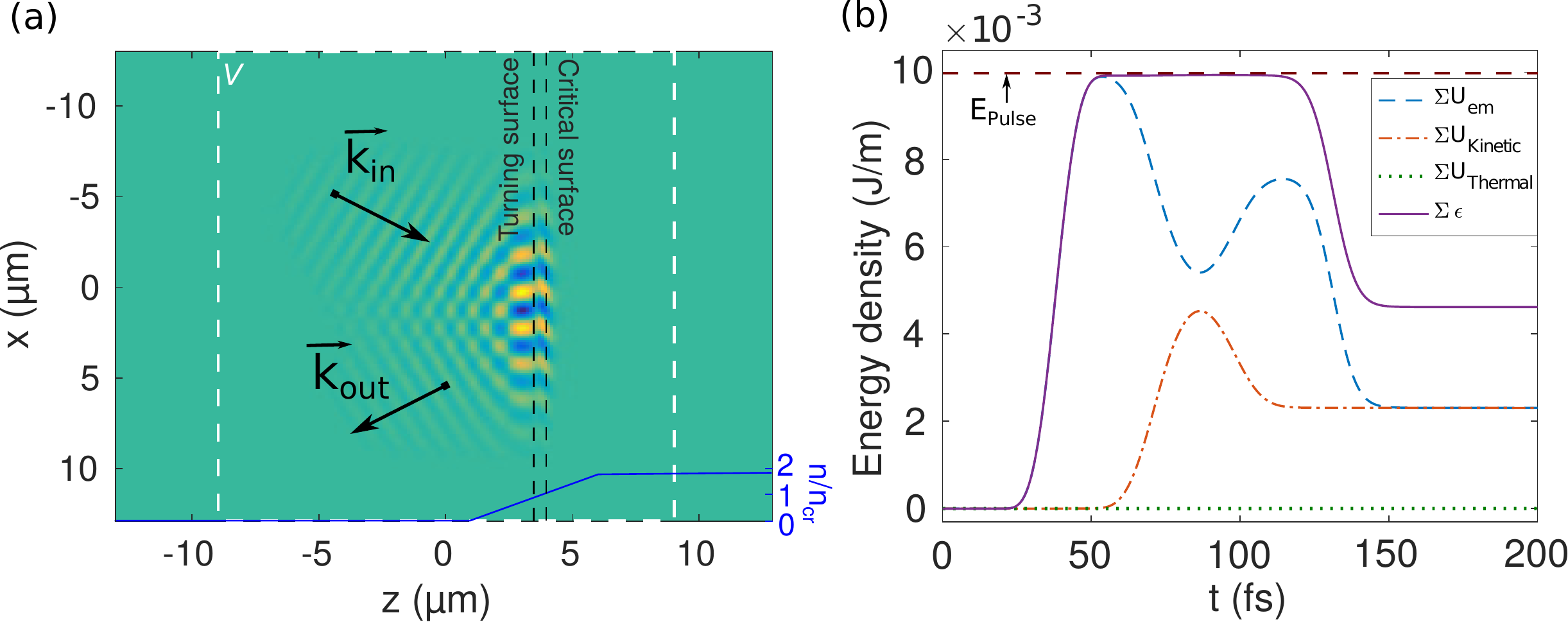}
\caption{(a) Concept of the simulation. A p-polarized laser field impinges on an inhomogeneous plasma slab (see main text). The blue line shows the plasma density profile. The white dashed rectangle shows the contour of the integration volume $V$. (b) Temporal evolution of the linear densities of: electromagnetic energy (dashed blue line), kinetic energy (dashed-dotted red line), thermal energy (dotted green line), and total energy (solid purple line) integrated over the volume $V$ for the laser pulse described in (a) with an incidence angle of 15$^\circ$ (see main text). The initial pulse linear density of energy $E_\mathrm{Pulse}$ is plotted as a dotted black line.}
\label{fig:Tm15}
\end{figure*}

Our simulation setup is shown in Fig.~\ref{fig:Tm15}(a). A spatially Gaussian laser beam with waist $w_\mathrm{0}=4~\mu$m is obliquely incident with an angle $\theta$ on an inhomogeneous plasma slab. The density profile is invariant in directions $x$, $y$ and linearly increases in $z$-direction. Its profile is shown as a solid blue line. The laser pulse is p-polarized and its amplitude is temporally described by $\sin^2\left(\pi\frac{t}{T} \right)$ for $t\leq T$ and is zero for $t> T$. We set $T=40$~fs in the amplitude profile which is equivalent to a Full Width at Half Maximum (FWHM) of 12.7~fs in the intensity profile. The beam is invariant along $y$-direction.
For reference, we display as black dashed lines the turning and critical surfaces at which the plasma density reaches respectively $n_{cr}\cos^2\theta$ and $n_{cr}$, which is the critical density.

We use the following initial densities and fields:
\begin{itemize}
\item The initial plasma densities $\rho_\mathrm{i}/m_\mathrm{i}$ and $\rho_\mathrm{e} /m_\mathrm{e}$ have the following profile (in cm$^{-3}$):

\begin{equation}
\left\lbrace
\begin{array}{ccc}
0 &  z<1\mu m\\
 0.57 \times 10^{21} (z-1) &  1\mu m\leq z \leq 6\mu m\\
2.85 \times 10^{21} & z > 6\mu m
\end{array}\right.
\end{equation}
 \noindent This profile corresponds to a plasma density scale length of $L=3.08~\mu$m, which is the length in $z$-direction at which the critical density is reached. A weak uniform background density of $10^{19}$~cm$^{-3}$ is added to avoid divisions by zero ({\it e.g.} when calculating velocity from momentum and density) and too strong discontinuity at the ramp onset.
\item Initial electric and magnetic fields are uniformly zero. The electromagnetic pulse is injected at $z=-12~\mu$m. The free-space wavelength is $\lambda = 0.8~\mu$m and the amplitude in free-space is $E_{0} = 10^{10}$~V/m. 
\end{itemize}

As we will further on investigate the conversion of energy between its different types in a finite volume, we define a volume $V$ - shown as white dashed line in Fig.~\ref{fig:Tm15}(a) - comprised between $z_1=-9~\mu$m and $z_2=9~\mu$m and which contains all the numerical points in $x$- and $y$-directions. The local electromagnetic energy density is \cite{griffiths_introduction_1999}:

\begin{equation}
U_\mathrm{em} = \frac{1}{2} \left[ \epsilon_\mathrm{0} ||{\bf E}||^2 + \frac{1}{\mu_\mathrm{0}} ||{\bf B}||^2 \right] 
\end{equation}
 The linear density of electromagnetic energy  $\Sigma U_\mathrm{em}$ is calculated by integrating the energy density $U_\mathrm{em}$ over the volume $V$ and dividing by the box thickness in $y$-direction.
 
We also define the thermal energy density as $\sum_\mathrm{k}  \frac{p_\mathrm{k}}{\gamma -1 }$ and the kinetic energy density as $\sum_\mathrm{k} \frac{1}{2}\rho_\mathrm{k} ||{\bf u_\mathrm{k}}||^2$. The corresponding linear density of thermal energy and kinetic energy are respectively written $\Sigma U_\mathrm{Thermal}$ and $\Sigma U_\mathrm{Kinetic}$. The linear density of total energy is  $\Sigma \epsilon = \Sigma U_\mathrm{em} + \Sigma U_\mathrm{Thermal}+ \Sigma U_\mathrm{Kinetic}$.

In Fig.~\ref{fig:Tm15}(b) we show the result of the simulation for an incidence angle $\theta = 15^\circ$. We plot the different types of energy defined above as a function of time. 
The linear density of energy of the input pulse in the $x-z$ plane is given by : $E_{Pulse} = \frac{E_\mathrm{0}^2}{2} \sqrt{\frac{\epsilon_\mathrm{0}}{\mu_\mathrm{0}}} \frac{3}{8} T w_0 \sqrt{\frac{\pi}{2}} = 10$~mJ/m and is shown as a black dotted line.

 The electromagnetic energy increases as the pulse enters into the volume $V$ between $t=0$ and $t<45$~fs. At $t=45$~fs, the pulse is completely contained in the volume and has not yet interacted with the plasma ramp. We see that the pulse energy corresponds to the predicted value of $10$~mJ/m.
  In the temporal window 45-130~fs, energy exchange with the plasma takes place (the kinetic energy of the plasma increases). Then, between $130$~fs and $160$~fs, the reflected pulse leaves the integration volume and the electromagnetic energy decreases. We see that the a fraction of energy remains in the volume. This residual energy corresponds to the one transferred from the electromagnetic wave into the electron plasma wave. The conversion factor for this simulation is $46\%$. Since the collisions frequency is taken to zero, the thermal energy linear density stays zero (dashed green line). We also see in the same graph that the total numerically integrated energy density, shown as solid purple line, actually corresponds to the incident pulse energy when the pulse is fully inside the volume $V$. This shows that our algorithm preserves energy.

\begin{figure}[h]
\centering
\includegraphics[scale=0.75]{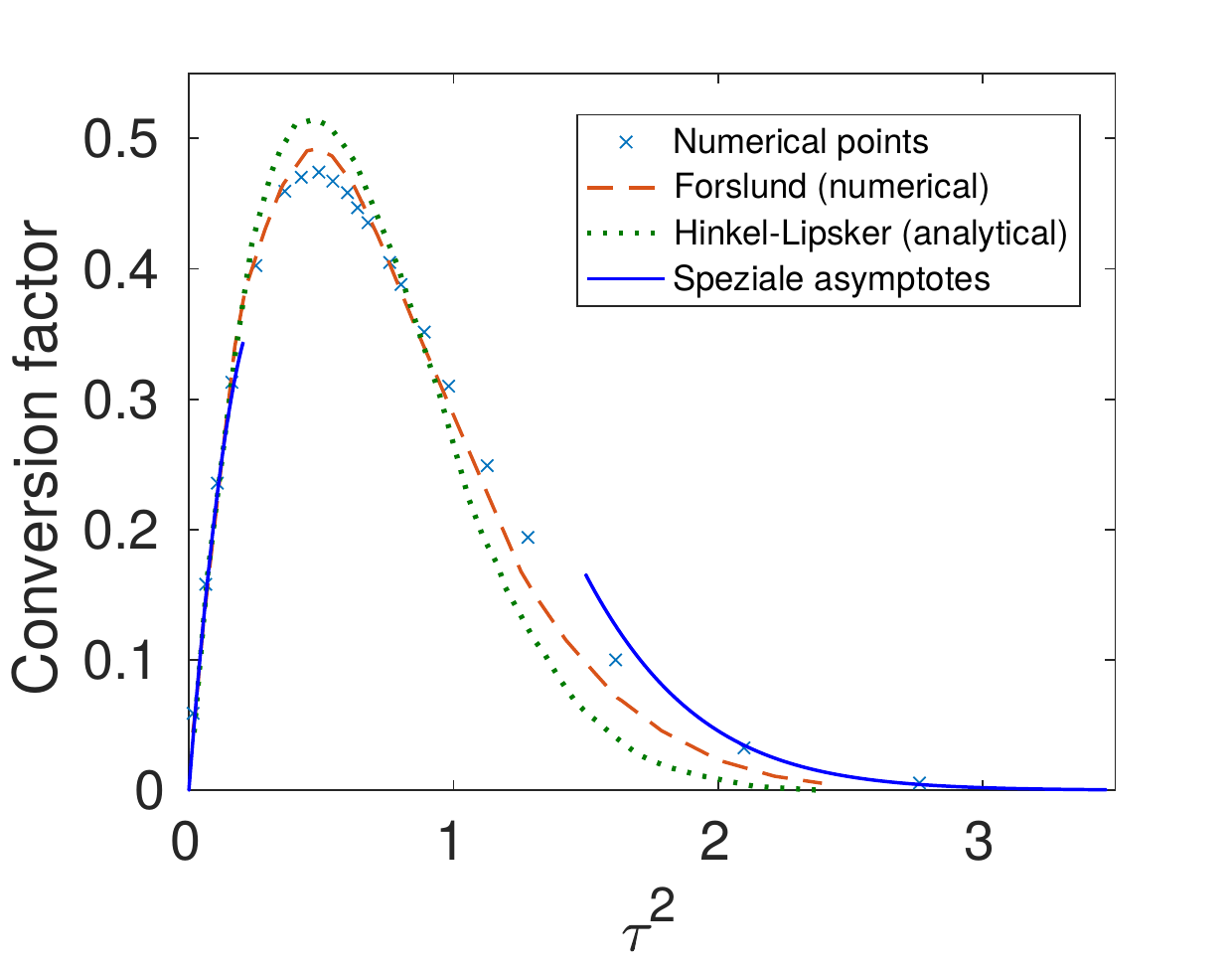}
\caption{Mode conversion factor as a function of $\tau^2 = \left(\frac{2 \pi L}{\lambda}\right)^{2/3}\sin^2\theta$. Blue crosses are our numerical results. Results from other references are shown for comparison: Forslund PIC simulations (dashed orange line) \cite{forslund_theory_1975}, Hinkel-Lipsker analytical model (dotted green line) \cite{hinkel-lipsker_analytic_1989} and Speziale asymptotes (solid blue lines) \cite{speziale_linear_1977}. }
\label{fig:Res}
\end{figure}

We repeat the same numerical simulation  with different angles of incidence. The conversion factor is shown in Fig.~\ref{fig:Res} as a function of $\tau^2 = \left(\frac{2 \pi L}{\lambda}\right)^{2/3}\sin^2\theta$. The results of the other references are shown for comparison: the theoretical asymptotes of Speziale (solid blue) \cite{speziale_linear_1977} {\it et al.}, the analytical results from Hinkel-Lipsker {\it et al.} (green dotted) \cite{hinkel-lipsker_analytic_1989} and the numerical results of Forslund {\it et al.} (orange dashed) \cite{forslund_theory_1975} (the latter were obtained for a relatively cold plasma  with $k_\mathrm{B} T_\mathrm{e} / (m_\mathrm{e} c^2) = 0.005$). We also note that another G. J. Pert\cite{pert_analytic_1978} obtains results comparable to those of Forslund {\it et al.}\cite{forslund_theory_1975}.
Our numerical results, shown as blue crosses, are in excellent agreement with all these previous results.

 \begin{figure*}
\centering
\includegraphics[scale=0.62]{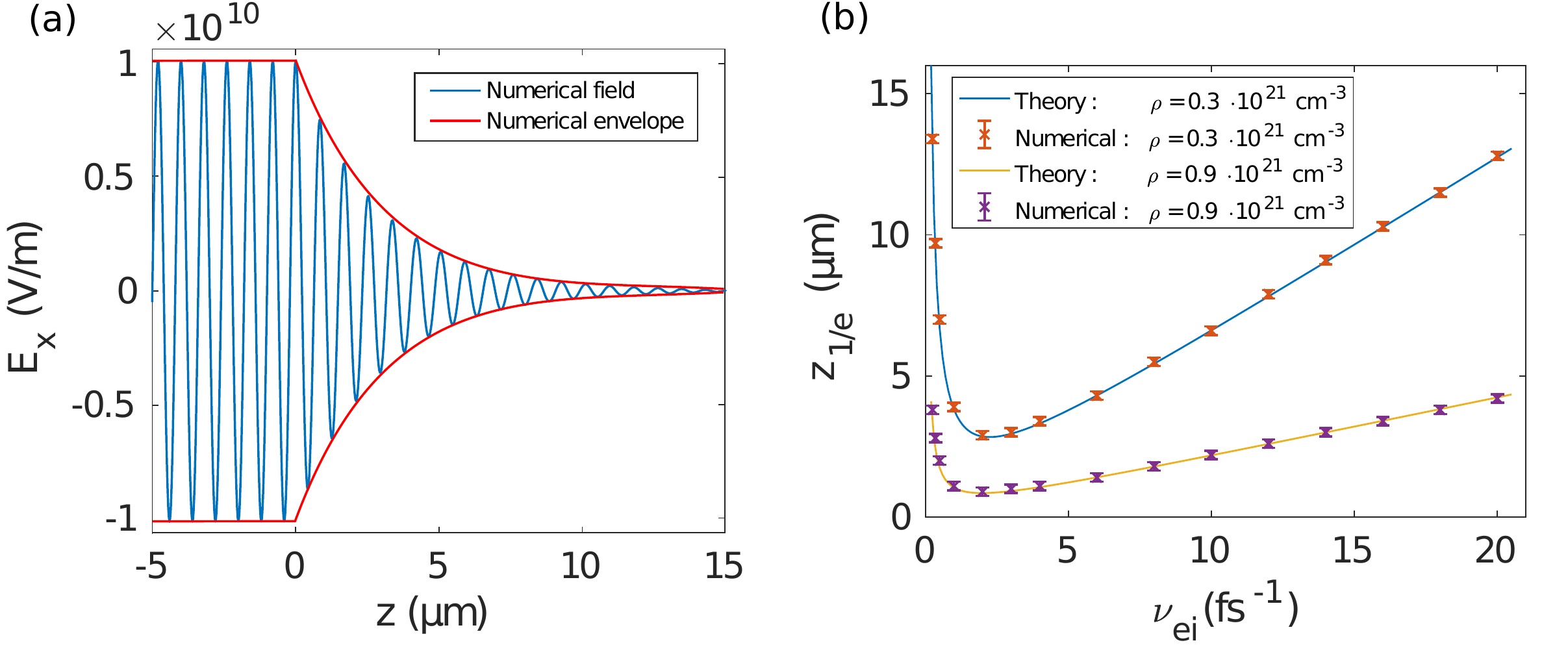}
\caption{(a) Spatial profile of electromagnetic wave amplitude propagating into a homogeneous collisional plasma of density $0.3\times 10^{21}$~cm$^{-3}$ and a collision time of 2~fs$^{-1}$. The vacuum/plasma interface is at $z=0~\mu$m. (b) Numerical and theoretical characteristic decay length $z_\mathrm{1/e}$ as function of collision frequency $\nu_\mathrm{ei}$. Results are plotted for two different densities:  $0.3 \times 10^{21}$~cm$^{-3}$ (blue line, orange crosses) and $ 0.9 \times 10^{21}$~cm$^{-3}$ (yellow line, purple crosses).}
\label{fig:t51}
\end{figure*}

\subsection{Test 4: Collisional plasmas}

Here we test the integration of the BGK collisional model. 
The Drude model for a cold uniform plasma of heavy ions provides the complex permittivity \cite{ashcroft_solid_2011}:
\begin{equation}
\epsilon = 1 - \frac{\omega_\mathrm{pe}^2 \tau_\mathrm{c}^2}{1 + \omega^2 \tau_\mathrm{c}^2} + i\frac{\omega_\mathrm{pe}^2 \tau_\mathrm{c}}{\omega (1 + \omega^2 \tau_\mathrm{c}^2)}
\label{eq:epim}
\end{equation}

We link the collision time $\tau_\mathrm{c}$ of Drude model to the electron-ion collision frequency of BGK model via $\tau_\mathrm{c} = \frac{1}{\nu_\mathrm{ei}}$. 
We check our model through the collisional damping of a monochromatic plane wave impinging on a plasma slab of uniform density. The distance at which the wave amplitude is reduced by a factor of 1/e is given by:
\begin{equation}
z_{1/e} = \frac{c}{\Im (\sqrt{\epsilon})\omega }
\label{eq:1oe2} 
 \end{equation}
 
Fig.~\ref{fig:t51} shows the results obtained for the following initial conditions:
\begin{itemize}
\item Initial plasma density is uniform and set to  $0.3 \times 10^{21}$~cm$^{-3}$ and to $0.9 \times 10^{21}$~cm$^{-3}$ in a second series of simulations. The plasma is only in region $z\geq~0\mu$m. 
\item The collision frequency $\nu_\mathrm{ei}$ is varied in the range 0.25-20~fs$^{-1}$.
\item The electric and magnetic fields are initially set to zero. A monochromatic electromagnetic plane wave propagating along the $z$-axis is progressively branched with normal incidence on the plasma. It is polarized along the $x$-axis with $\lambda = 0.8~\mu$m and amplitude of $E_{0} = 10^{10}$~V/m. 
\end{itemize}

In Fig.~\ref{fig:t51}(a), we plot the amplitude of the electric field as a function of propagation distance $z$. The exponential damping is observed as expected. In Fig.~\ref{fig:t51}(b), we report the values of the decay length $z_\mathrm{1/e}$ which we compare to the analytical results from Eq.~(\ref{eq:1oe2}). The error bars correspond to the graphical measurement. An excellent agreement is found. In parallel, we measured the evolution of wavelength in same conditions, which we compared with the theoretical value $2\pi c / \left[ \omega \Re (\sqrt{ \epsilon} ) \right]$ (not shown). The discrepancy between numerical and theoretical results were less than 1\%.

\begin{figure}[h]
\centering
\includegraphics[scale=0.75]{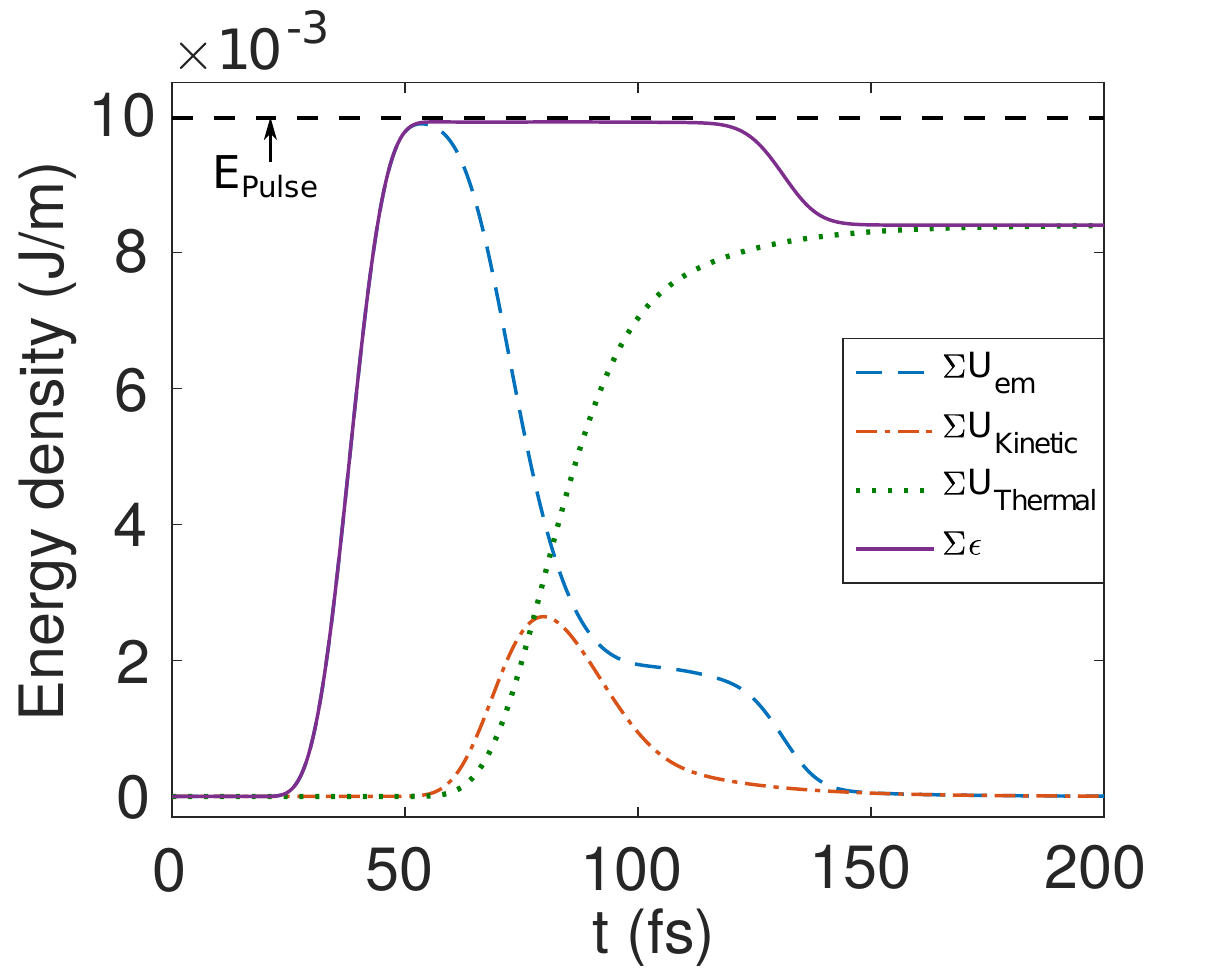}
\caption{
Temporal evolution of the linear densities of: electromagnetic energy (dashed blue line), kinetic energy (dashed-dotted red line), thermal energy  (dotted green line), and total energy (solid purple line) integrated over the volume $V$ for the laser pulse described in section \ref{sec:waveConversion} and a collision frequency $\nu_\mathrm{ei} = 0.05$~fs$^{-1}$. The initial pulse linear density of energy $E_\mathrm{Pulse}$ is plotted as a dashed black line.}
\label{fig:colllener}
\end{figure}

Separately, we also test the conservation of energy within the collisional plasma. The same simulation as for Fig.~\ref{fig:Tm15}(b) is performed in the collisional plasma with a collision frequency $\nu_\mathrm{ei} = 0.05$~fs$^{-1}$ (see Fig.~\ref{fig:colllener}). We observe a similar evolution as in section \ref{sec:waveConversion}, except that now, due to the collisions, the energy of waves is converted to thermal energy with time. The total density of energy is preserved as one can see from the purple line in the temporal window [45-130]~fs, when the pulse is entirely in the integration volume $V$. It matches the initial pulse energy density.

Overall, these results validate the integration of the collisional model.

\subsection{Test 5: Wakefield generation}

Here, we test the algorithm in a nonlinear regime, where transport effects are non-negligible. For this, we consider the interference between two contrapropagative laser pulses in a sub-critical plasma. The interference generates a periodic pattern of low and high intensities along the beams axis. The ponderomotive force swiftly pushes the electrons away from high-intensity regions \cite{Mulser2010}, which creates a net space charge and consequently a wakefield. The use of a standing wave therefore allows the observation of strong ponderomotive effects without using relativistic laser pulses \cite{Smorenburg_2011}.

We reproduce this physical situation in this test. We consider the following initial conditions:
\begin{itemize}
\item The sub-critical plasma has initial uniform densities $\rho_\mathrm{i}/m_\mathrm{i} = \rho_\mathrm{e} /m_\mathrm{e} = 5\times 10^{19}$cm$~^{-3}$ and the collision frequency is $\nu_{ei}=0.5$~fs$^{-1}$.
\item The electric and magnetic fields are initially set to zero in the plasma. Two incident electromagnetic pulses are injected to generate a standing plane wave. The first one is injected in $z=-20\mu$m and is propagating along $z$-direction. The second one is injected in $z=20\mu$m and is propagating in the opposite direction. Both of them are polarized along  $x$ axis with $\lambda = 0.8~\mu$m and amplitude of $E_{0} = 1\times 10^{11}$~V/m. The intensity profile of both pulses is a $(\sin^2)^2$ with FWHM 47.6~fs.
\end{itemize}

\begin{figure}
\centering
\includegraphics[scale=0.75]{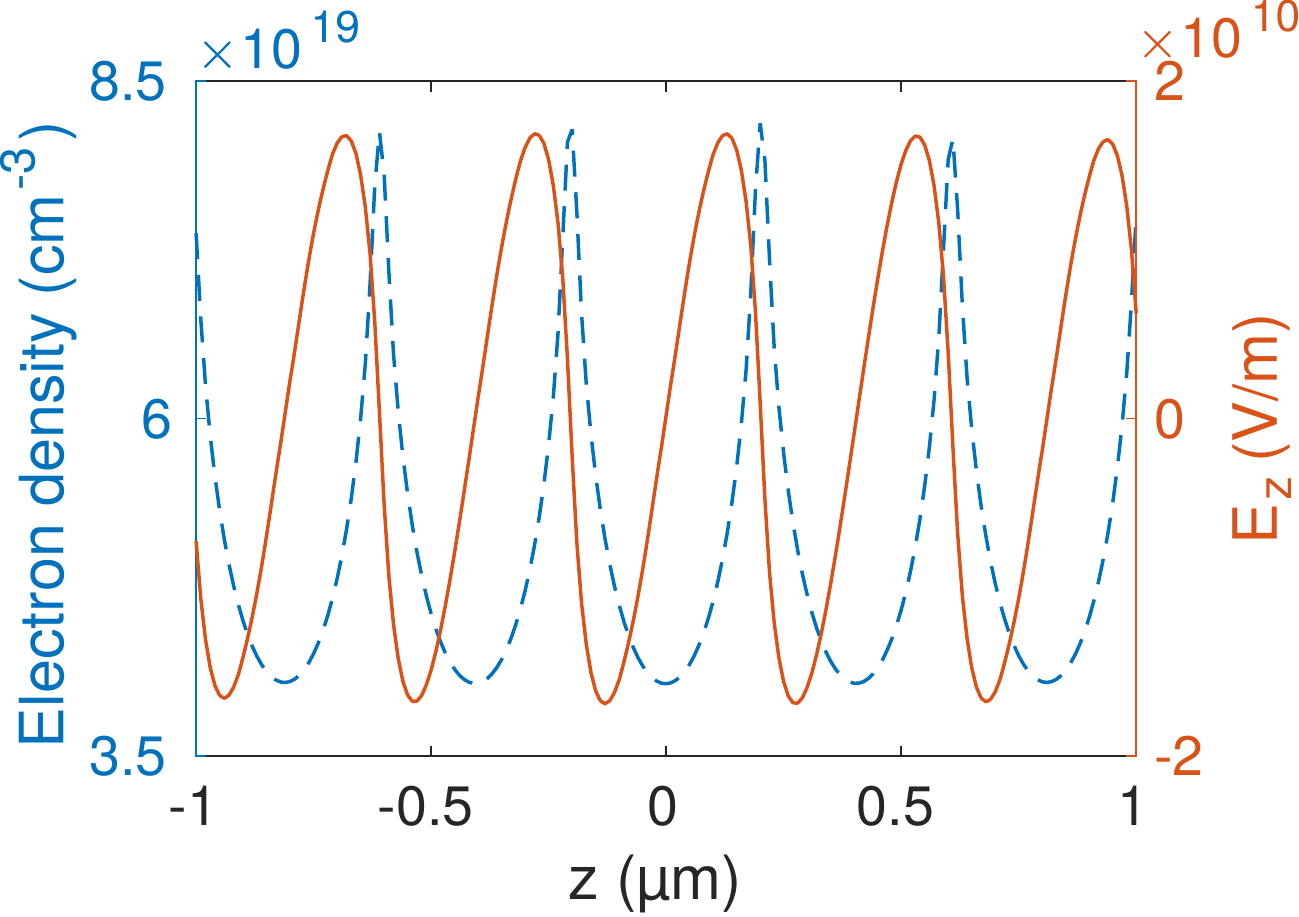}
\caption{Electron density profile (dashed blue line) and wakefield $E_\mathrm{z}$  (red line) at the time where the two interfering pulses cross at $z = 0$.}
\label{fig:pon}
\end{figure}

The simulation is run until the pulses peaks cross at $z = 0$.
Fig.~\ref{fig:pon} shows, for that time, the electron density profile (dashed blue line) and the longitudinal component of the electric field $E_\mathrm{z}$ as a function of the propagation distance $z$. 

Our results fit qualitatively very well with reference \cite{gibbon_short_2005}. On a quantitative point of view, reference P.-W Smorenburg \textit{et al.}\cite{Smorenburg_2011} provide an estimate of wakefields of $\sim 10^9$~V/m for standing wave field strength of the order of $\sim 10^{15}$~W/cm$^{2}$. Thus, we performed the same simulation as for Fig.~\ref{fig:pon} but in a collionless plasma and with a field which corresponds to a standing wave with an intensity of $ 10^{15}$~W/cm$^{2}$. We obtained a wakefield $E_\mathrm{z} = 5 \times 10^8$~V/m. Our results are therefore in good quantitative agreement with Ref.~\cite{Smorenburg_2011}. In addition, we note that very strong modifications of the density profile (up to 50\%) are reached without compromising the stability of the simulation.

\section{Conclusions}
In conclusion, we have developed a simple two-fluid plasma model solver. Our algorithm is composed of a PSTD solver for Maxwell's curl equations and of a combination of LWLFn and RK4 algorithms via a Strang splitting for the fluid equations. We have also integrated a simple collisional model. We have validated the solver against several well-known problems of laser-plasma interaction. In all cases, very good agreement has been achieved. 

This solver is simple and robust. Its implementation is relatively straightforward since, in contrast with complexities contained in most of the two-fluid plasma solvers, it relies only on finite difference schemes and FFT. This type of solver is a good compromise for physical problems that do not involve strong discontinuities. It can represent a simple first step before investing time and efforts into more elaborate solvers. We anticipate that this approach will open new perspectives in the fields of nonlinear plasmonics and laser-plasma interaction in solid.

\section*{Acknowledgments}
Numerical simulations have been performed using the M\'{e}socentre de Calcul de Franche-Comt\'{e}. The research leading to these results has received funding from the European Research Council (ERC) under the European Union's Horizon 2020 research and innovation program (grant agreement No 682032-PULSAR), R\'egion Bourgogne Franche-Comt\'e and the EIPHI Graduate School (ANR-17-EURE-0002). 

\bibliography{Biblio}

\end{document}